\documentclass[aps,pra,twocolumn,superscriptaddress]{revtex4}
\usepackage{amssymb}
\usepackage{graphicx}
\usepackage{amsmath}
\usepackage{color}
\usepackage{bbm}
\usepackage{multirow}
\usepackage{hhline}

\begin{document}

\title{The Single Particle Excitation Spectrum of Degenerate Fermi gases in a ring cavity}
\author{Chao Feng}
\affiliation{Institute for Advanced Study, Tsinghua University, Beijing, 100084, P. R. China}
\author{Yu Chen}
\email{yuchen.physics@cnu.edu.cn}
\affiliation{Center for Theoretical Physics, Department of Physics, Capital Normal University, Beijing, 100048, P. R. China }

\begin{abstract}
By considering a spin-$\frac{1}{2}$ degenerate Fermi gases in a ring cavity where strong interaction between atoms and light gives rise to a superradiance, we find the cavity dissipation could cause a severe broadening in some special cases, breaking down the quasi-particle picture which was constantly assumed in mean field theory studies. This broadening happens when the band gap resonant with polariton excitation energy. Interestingly enough, this broadening is highly spin selective depending on how the fermions are filled and the spectrum becomes asymmetric due to dissipation. Further, a non-monotonous dependence of the maximal broadening of the spectrum against cavity decay rate $\kappa$ is found and the largest broadening emerges at $\kappa$ comparable to recoil energy.
\end{abstract}

\maketitle
\section{Introduction}
Recent developments in experiments extend traditional cavity QED\cite{cavityQED1,cavityQED2} to a many body system by putting quantum degenerate gases into cavities \cite{Esslinger07,Colombe,Esslinger10}. Strong interactions between atoms and light are realized, and a steady state superradiance is achieved\cite{Dicke, Esslinger10}. This field naturally combines the studies of the non-equilibrium driven-dissipative systems and the quantum many body systems together\cite{EsslingerRev13}. As coherence is very important for realizing superradiance, therefore how dissipation influences this many body state is of fundamental importance.  Regarding nontrivial effect of dissipation, one example is the dynamical critical exponent being shifted by the presence of dissipation. A theoretical prediction of the critical exponent shift is from $0.5$ to $1$ \cite{Domokos11,Sachdev13a}. While the measurement on static and dynamical structure factor shows the flux critical exponent is shifted to 0.7 in normal phase side, and 1.1 in superradiant side\cite{Roton, DissipationDicke,DynStruct}. The mechanism for the critical exponent shift is not yet fully understood. Another example is, in a cavity superradiance assist hopping system, dissipation could induce an emergent electric field and drives atoms to flow in one direction, forming a ``real space Fermi sea" for degenerate fermions\cite{ZhengWei16}.

Although there are nontrivial effects for dissipation, there are still many cases these effect are not major and can be neglected. There are many previous theoretical works\cite{cavityBEC4, cavityFermion1, cavityFermion2, cavityFermion3, cavityFermion4, TSRExpFermion, Simons09, Simons10, Ciuti13, He13, Hemmerich15, Yu16, Brennecke16, Brennecke15} on superadiance in a cavity assume that the atomic excitations are well defined quasi-particles, so that dissipation effect could be neglected. However, the validity of this  quasi-particle assumption has not yet been justified in many cases. Some efforts are devoted to calculating dynamics and dissipation of Bose gases in a cavity\cite{Simons10a,Simons12}, where the spectrum broadening is found and the critical behavior of the quasiparticle lifetime is analyzed\cite{Domokos11,Sachdev13a}. Even more, extra spectrum broadening by Baeliev damping is predicted for interacting bosons\cite{Domokos14}. Discussion of steady state distribution of fermions in a cavity can be found in Ref.\cite{Pizza13}, but studies for fermion spectrum is still quite limited. In this article, we will study the fermion excitation spectrum by a systematic Keldysh formalism which could be applied to general nonequilibrium many body systems. To keep our study as simple as possible, we choose to study one dimensional spin-$\frac{1}{2}$ Fermi gases in a ring cavity. Since absorption of cavity photon will flip spin together with a momentum transfer, therefore the condensation of cavity field will induce a spin-orbit-coupling (SOC) of fermions. This is originally proposed by Han Pu and his coworkers\cite{Pu14,Pu15} in a bosonic system.


In this article, we will first introduce the experimental setup and corresponding hamiltonian in section II, then we will establish the Keldysh formalism for calculating fermion's excitation spectrum and distribution function under steady state assumption in section III. Dyson equations for cavity field self-energies and fermion self-energies are given. In section IV, we will present the numerical results for cavity photon spectrum and fermion excitation spectrum in different parameter regions. A spin selection of spectrum broadening is discovered to be connected with the fermion occupation. Maximal spectrum broadening is found to be non-monotonous dependent on cavity decay rate $\kappa$. For small $\kappa$ and large $\kappa$, the spectrum is more close to well defined quasi-particle spectrum, and the largest broadening happens for moderate $\kappa$ around recoil energy. This result is in accord with how fast a steady state could be reached in a one-dimensional lattice fermion system\cite{CavityChiralSFermion}, where steady state is most difficult to be reached for $\kappa\sim E_r$. In this parameter region, photon loss rate resonant with atom motion, therefore the adiabatic approximation breaks down. Both calculations show that $\kappa\sim E_r$ is the most unstable case. Finally, we make conclusion in section V. 

\section{Set Up}
\begin{figure}[t]
\includegraphics[width=2.6in]{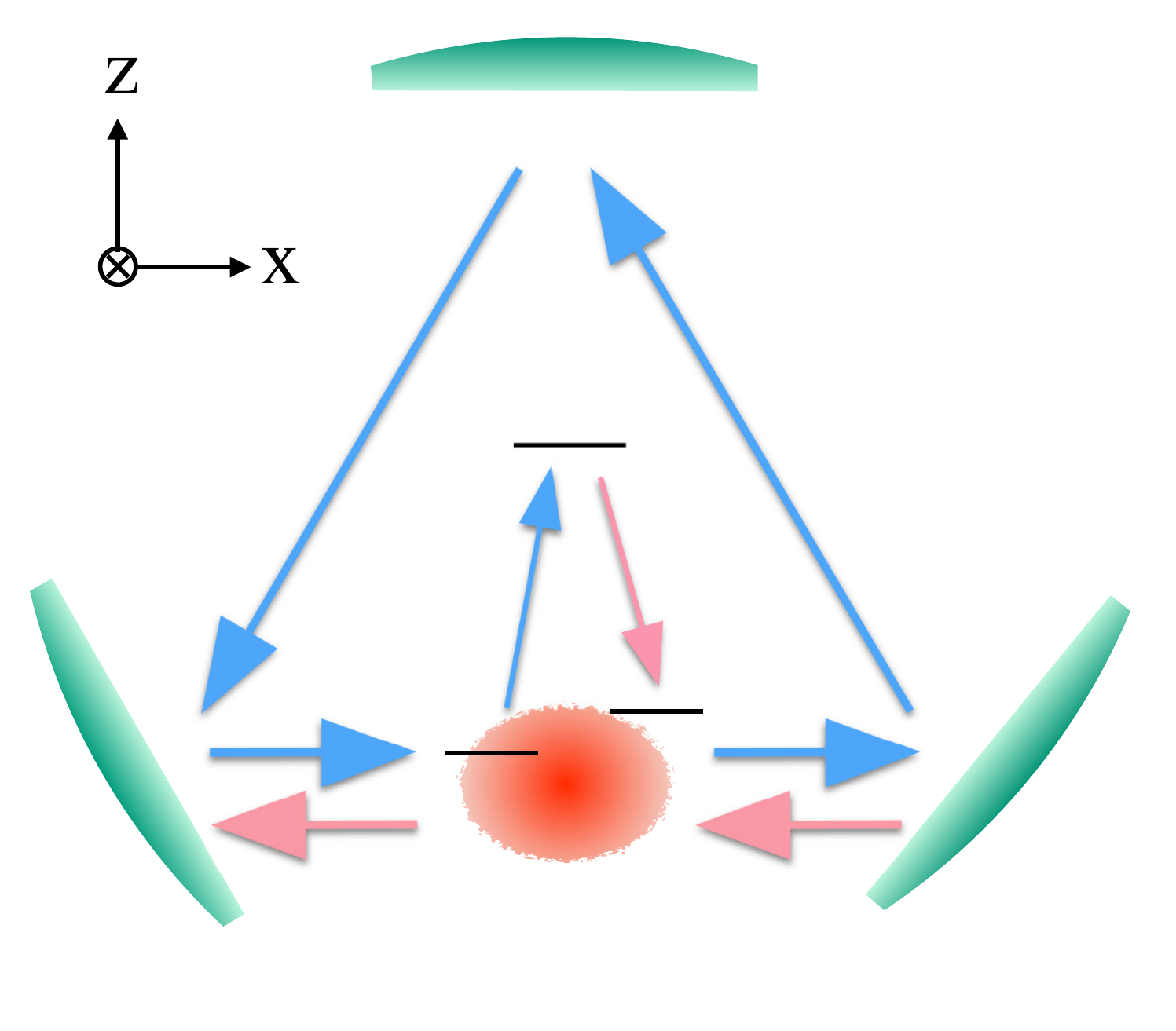}
\caption{Experimental setup for cavity induced spin-orbit-coupling in spin-$1/2$ Fermi gases. Pumping field is labeled in blue while the cavity field is labeled in red. }\label{Set}
\end{figure}

Here we propose to put degenerate spin-$\frac{1}{2}$ Fermi gases into a ring cavity as shown in Fig.~\ref{Set}. Here we suppose the interactions between fermions could be neglected. The ring cavity supports two traveling wave modes in clockwise and anti-clockwise directions. Here we suppose the pumping laser is in anti-clockwise direction polarized in $\hat{y}$ while a one dimensional fermions cloud is laid in $\hat{x}$ direction. Photons interact with the atoms by a two photon Raman process, in which spin is flipped together with a momentum transfer $2k_0$. The cavity mode is clockwise and polarized in $\hat{z}$ direction. The hamiltonian of the system could be then formulated as 
\begin{eqnarray}
\hat{H}&=&\int dx \hat{\Psi}^\dag(x)\hat{H}_{\rm at}\hat{\Psi}(x)-\tilde{\delta}_c\hat{a}^\dag\hat{a}+i\epsilon_p(\hat{a}^\dag-\hat{a}),\\
\hat{H}_{\rm at}\!\!&=&\frac{(p_x+k_0\sigma_z)^2}{2m}+\eta(\hat{a}\sigma^++\hat{a}^\dag\sigma^-)-\mu,
\end{eqnarray}
under rotating wave approximation (RWA).
The atomic field operator $\hat{\Psi}(x)$ could be written as $\hat{\Psi}(x)=(1/L)\sum_{k}(\hat{c}_{k\uparrow},\hat{c}_{k\downarrow})^T e^{ikx}$ in second quantization form, $L$ is the size of atom cloud in x direction. $\sigma^\pm=\sigma_x\pm i\sigma_y$ are spin ladder operators, where $\sigma_{i=x,y,z}$ are Pauli matrices. In original frame the Raman process will flip the atom's spin at the same time shift the momentum by $k_0{\bf e}_x$, here $\hat{H}_{\rm at}$ is the atomic hamiltonian after unitary transformation $\hat{U}=e^{-ik_0x\sigma_z}$. It is clear the hamiltonian is invariant under a U(1) transformation $\hat{a}\rightarrow \hat{a} e^{i\phi}$, $\hat{c}_{k\sigma}\rightarrow \hat{c}_{k\sigma}e^{i\sigma \phi/2}$ if we turn off $\epsilon_p$ term. Therefore the hamiltonian without $\epsilon_p$ term has $U(1)$ symmetry.  
$\epsilon_p$ term is an explicit symmetry broken term coming from direct exchange of photons between cavity and pumping laser field. $\tilde{\delta}_c=\omega_{\rm p}-\omega_{\rm c}$ is the cavity detune frequency. $\omega_{\rm c}$ and $\omega_{\rm p}$ are frequency for cavity mode and pumping field frequency respectively. $\eta=\Omega_{\rm p} g_0/\Delta_a$ is the atom cavity coupling strength, with $\Omega$ being the pumping laser amplitude, $g_0$ being single atom cavity coupling strength, $\Delta_a$ being ac Stark shift of the atom. The ring cavity's decay rate is assumed as $\kappa$.

When $\epsilon_p\neq 0$ is turning on, the U(1) symmetry will be explicitly broken so that the vacuum expectation value of $\hat{a}$ is nonzero, leading to a SOC in fermions. We stress that this SOC by cavity photon condensation is in general different from previous SOC by Raman lasers \cite{Spielman13,Pu14R,Hui15} where the atomic spectrum cannot fluctuate.

\section{Method}

\subsection{Keldysh formalism}
Here our aim is to obtain the fermionic excitation spectrum in presence of dissipation. Since this dissipative system does not in general reach thermal equilibrium in long time limit, therefore we choose Keldysh formalism to formulate this problem\cite{Kamenev}. In ordinary equilibrium field theory we assume the density operator $\hat{\rho}=\exp(-\beta\hat{H})$, and all correlation functions are thermal ensemble averaged functions. However, in general nonequilibrium systems the distribution function is unkown and varying with time. Keldysh formalism does not assume any \emph {ad hoc} distribution function and both the distribution function and single particle excitation spectrum could be solved at the same time.

\begin{figure}[h]
\includegraphics[width=8.5cm]{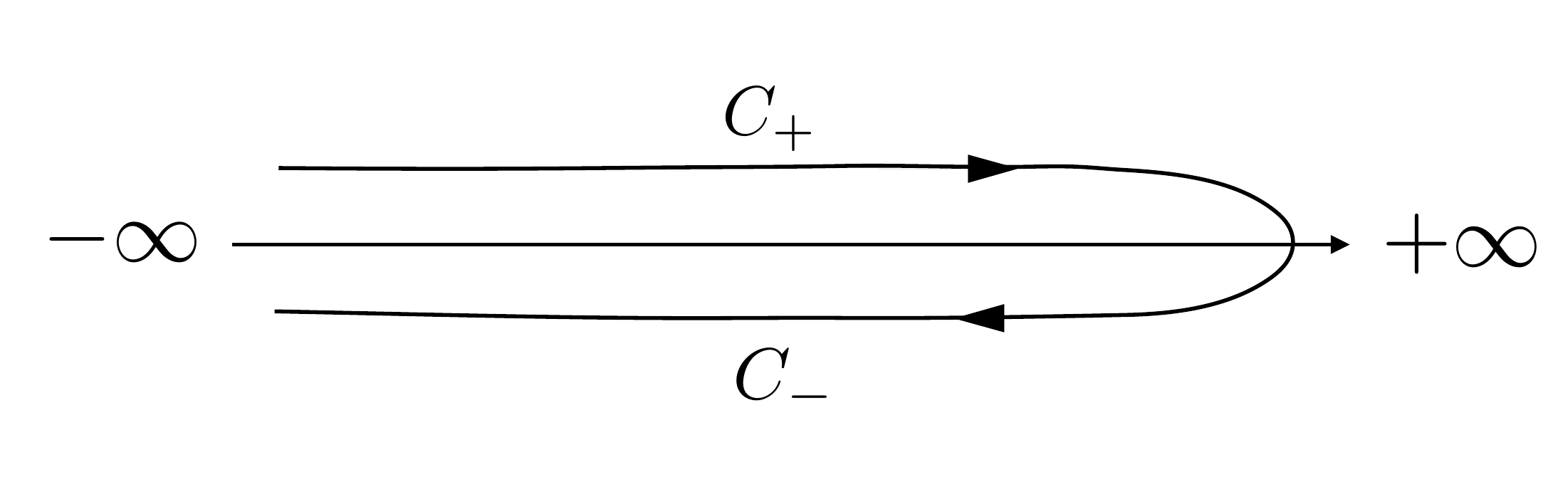}
\caption{Illustration of closed time path.}\label{CTPP}
\end{figure}
According to our setup, free fermions are coupled to a single mode cavity whose decay rate is $\kappa$. The action could be written down by a closed time path integral as
\begin{eqnarray}
S\!&=&\!\!\sum_{\lambda=\pm}\int_{C_\lambda} \!\!\!\!dt(-)^{\lambda}\!\left[\frac{}{}\!\!\left(\bar{a}_\lambda(\!-i\partial_t\!+\!\delta_c\!+\!i\lambda\kappa) a_\lambda\!+\!i\epsilon_p(\bar{a}_\lambda\!-\!a_\lambda)\right)\!+\!\right.\nonumber\\
&&\!\!\left.\!\!\!\!\int\!dx\bar{\Psi}_\lambda\!\left(\!-i\partial_t\!-\!\frac{(i\partial_x\!-\!k_0\sigma_z)^2}{2m}\!+\!\eta(\bar{a}_\lambda\sigma^-\!\!+\!a_\lambda\sigma^+)\!\right)\!\Psi_\lambda\!\right],
\end{eqnarray}
where $C_{\lambda=\pm}$ are two branches of closed time path, $C_+$ is the positive branch and $C_-$ is the negative branch, as is shown in Fig.\ref{CTPP}. Cavity field $a_\lambda(t)$ and atom field $\Psi_\lambda(x,t)$ are defined on path $C_\lambda$ respectively. $\Psi_\lambda(x,t)=(\psi_{\lambda\uparrow},\psi_{\lambda\downarrow})^T$ is the fermion spinor field. After we introduce Keldysh rotation, we define $a_{\rm cl}\equiv(a_++a_-)/\sqrt{2}$, $a_{\rm q}\equiv(a_+-a_-)/\sqrt{2}$, and $\psi_{1\sigma}\equiv\frac{1}{\sqrt{2}}(\psi_{+\sigma}+\psi_{-\sigma})$, $\psi_{2\sigma}\equiv\frac{1}{\sqrt{2}}(\psi_{+\sigma}-\psi_{-\sigma})$, $\bar{\psi}_{1\sigma}\equiv\frac{1}{\sqrt{2}}(\bar{\psi}_{+\sigma}-\bar{\psi}_{-\sigma})$, $\bar{\psi}_{2\sigma}\equiv\frac{1}{\sqrt{2}}(\bar{\psi}_{+\sigma}-\bar{\psi}_{-\sigma})$ where $\sigma=\uparrow$ or $\downarrow$ is the spin index. Then the action can be expressed as
\begin{eqnarray}
S=S_c+S_{\rm at}+S_{\rm int}.
\end{eqnarray}
The action of cavity field is
\begin{eqnarray}
S_c=\frac{1}{2}\int_{-\infty}^\infty\frac{d\omega}{2\pi}A^\dag(\omega)\left(\begin{tabular}{cc}
$0$&$(\pi_{0}^{-1})^{A}$\\
$(\pi_{0}^{-1})^{R}$&$(\pi_{0}^{-1})^{K}$
\end{tabular}
\right)A(\omega),
\end{eqnarray}
$A(\Omega)=(a_{\rm cl}(\omega),a_{\rm cl}^*(-\omega),a_{\rm q}(\omega),a_{\rm q}^*(-\omega))^T$. The inverse of propagators are defined as
\begin{eqnarray}
(\pi_0^{-1})^R=\left((\pi_0^{-1})^A\right)^\dag=\left(
\begin{tabular}{cc}
$\omega+\delta_c+i\kappa$&$0$\\
$0$&$-\omega+\delta_c-i\kappa$
\end{tabular}
\right),
\end{eqnarray}
\begin{eqnarray}
(\pi_0^{-1})^K=\left(
\begin{tabular}{cc}
$2i\kappa$&$0$\\
$0$&$2i\kappa$
\end{tabular}
\right).
\end{eqnarray}
The atomic action is
\begin{eqnarray}
S_{\rm at}=\int drdr'\bar{\Psi}^T(r)
\left(\begin{tabular}{cc}
$L_0^R(r;r')$&$L_0^K(r;r')$\\
$0$&$L_0^A(r;r')$
\end{tabular}
\right)
\Psi(r'),
\end{eqnarray}
where $r=(x,t)$, $dr=dxdt$ and $\Psi(x,t)=(\psi_{1\uparrow},\psi_{1\downarrow},\psi_{2\uparrow},\psi_{2\downarrow})^T(x,t)$. $L_0^{R,A}(r,r')=L_0^{R,A}(r-r')$ and $L_0^K(r,r')=L_0^{K}(r-r')$ are inverse retard, advanced and Keldysh Green's functions. Their definition could be better expressed in frequency and momentum space, which is
\begin{eqnarray}
S_{\rm at}\!=\!\!\int \!\!\frac{d\epsilon dk}{(2\pi)^2}\bar{\Psi}^T\!
\left[\!\begin{tabular}{cc}
$\epsilon\!+\!i0^+\!-\!\frac{(k+k_0\sigma_z)^2}{2m}\!\!$&$2i0^+ F(\epsilon)$\\
$0$&$\!\!\!\epsilon\!-\!i0^+\!-\!\frac{(k+k_0\sigma_z)^2}{2m}\!$
\end{tabular}
\!\right]\!\Psi,
\end{eqnarray}
where $\Psi(k,\epsilon)$'s variables are neglected, $F(\epsilon)=1-2n_F(\epsilon)$ and $n_F(\epsilon)=(\exp((\epsilon-\mu)/T)+1)^{-1}$ is the zero temperatuare Fermi-Dirac distribution function. Finally the interaction action could be written as
\begin{eqnarray}
S_{\rm int}=\!\frac{\eta}{\sqrt{2}}\!\int\! dr \!\!\sum_{i=q,\rm cl}\bar{\Psi}^T(r)(a_i(t)\sigma^++\bar{a}_i(t)\sigma^-)\hat{\gamma}_i\Psi(r)
\end{eqnarray}
where $\hat{\gamma}_q=\sigma_x$,  $\hat{\gamma}_{\rm cl}={\mathbbm 1}$ are defined in Keldysh space. 
\begin{figure}[t]
\includegraphics[width=6cm]{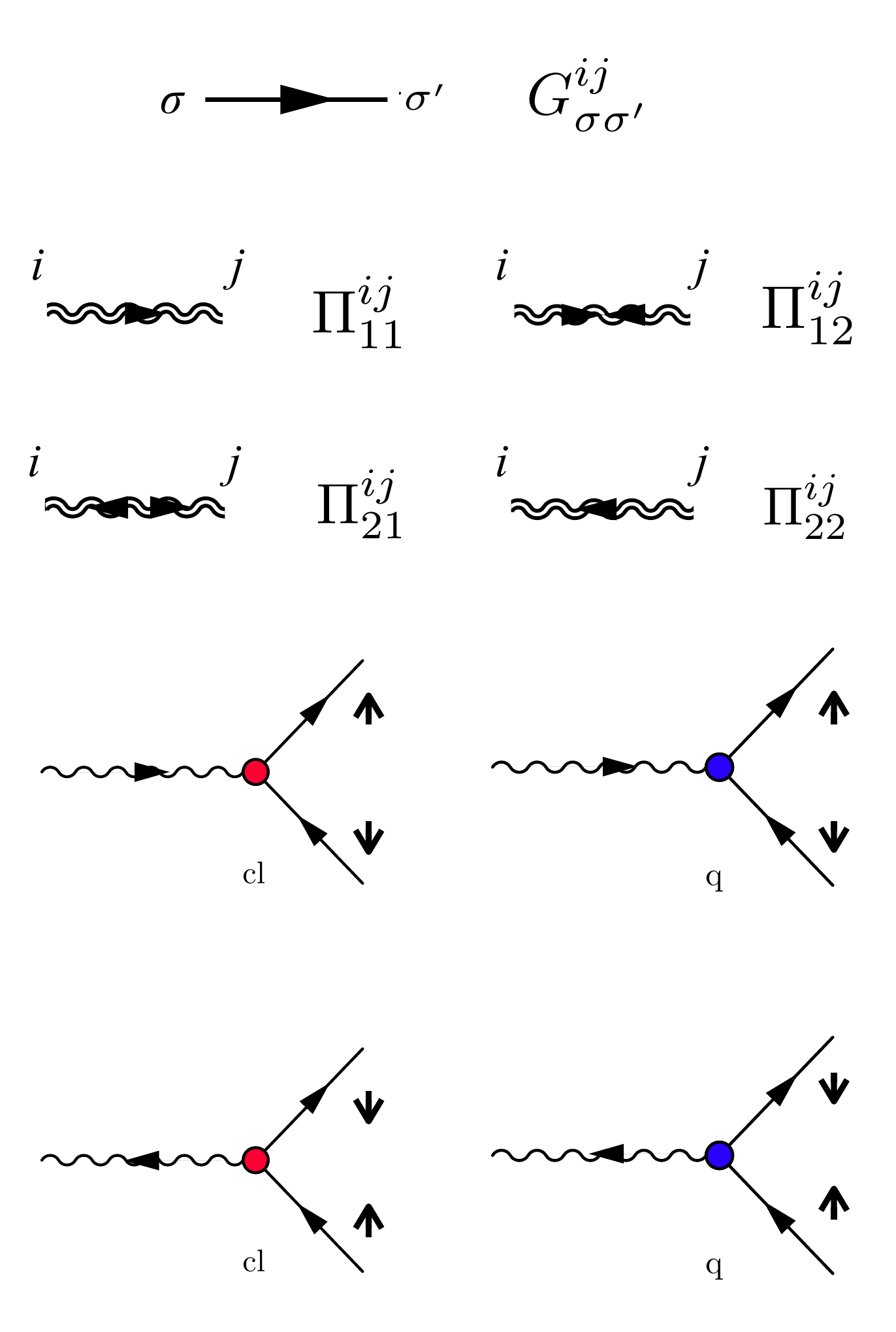}
\caption{Feynman diagrams for fermion propagators, cavity photon propagators and cavity-atom interaction. Fermion propagators are given by solid lines, representing $G^{ij}_{\sigma\sigma'}(k,\epsilon)$, where $\sigma$ and $\sigma'$ are spin indices. i and j are in Keldysh space, $G^{11}=G^R$, $G^{12}=G^K$, $G^{21}=0$ and $G^{22}=G^A$. Double wiggly lines are propagators of dressed cavity photon, labeled as  $\Pi^{ij}_{ab}$, where $i, j=q, {\rm cl}$ are in Keldysh space, $a, b$ are in T space. $\Pi^{\rm cl,cl}=\Pi^K$, $\Pi^{{\rm cl}, q}=\Pi^R$ and $\Pi^{q,\rm cl}=\Pi^A$, $\Pi^{q,q}=0$. The blue vertex means a matrix $\hat{\gamma}_q=\sigma_x$ in Keldysh space, and the red vertex means a matrix $\hat{\gamma}_{\rm cl}={\mathbbm{1}}$ in Keldysh space. The incoming arrow for photon is annihilation process of photon while outgoing arrow means creation of cavity photon. Spin flip process is directly shown in these Feynman rules. Feynman rules for these four vertices are $\eta(\hat{\gamma}_{\rm cl})_K\times (\sigma^+)_{\rm spin}$.  } \label{Feyn}
\end{figure}

In this strongly interacting atom light system, one basic question is whether there is a steady state solution in long time limit. This question could be answered as long as we know the exact form of Keldysh Green's function for cavity field $\hat{\Pi}^K(t,t')=-i\langle (a_{\rm cl}(t), \bar{a}_{\rm cl}(t))^T(\bar{a}_{\rm cl}(t'), a_{\rm cl}(t'))\rangle$, and Keldysh function of fermions $G^K(r,r')=-i\langle \Psi_1(r)\bar{\Psi}_2(r')\rangle$. These functions are related to the excitation spectrum and the distribution function evolution. Without assuming that the system being time translational invariant, $\Pi^K(t,t')$ will be not only a function of time difference $t-t'$, but also a function of $(t+t')/2$, which manifests time evolution of distribution function. If for large $(t+t')/2$, $\Pi(t,t')$ becomes  $(t+t')/2$ independent, then we say the system reaches steady state. In present work we will focus on the spectrum study, therefore we assume that steady state could be always reached in long time limit.

By this assumption, all Green's functions become functions of $t-t'$ only. In steady state approximation, we assume the cavity expectation value becomes $\langle a\rangle=\alpha$ in long time limit, where $\langle\cdot\rangle$ is ground state average. We define retard and advanced Green's function for fermions are $G^R(r,r')=-i\theta(t-t')\langle \Psi_1(r)\Psi^\dag_1(r')\rangle$, $G^A(r,r')=-i\theta(t'-t)\langle\Psi_2(r)\Psi^\dag_2(r')\rangle$. Let us assume the fermions are in zero temperature, with the cavity vacuum expectation value $\alpha$ self consistently determined by steady state equaion, then the zeroth order Green's function could be written as 
\begin{eqnarray}
G^{0,R}_{\uparrow\uparrow}(k,\epsilon)&=&\frac{1}{2\Delta_k}\!\left(\frac{\Delta_k+v_0k}{\epsilon\!+\!i0^+-\xi_k^+}+\frac{\Delta_k-v_0k}{\epsilon\!+\!i0^+-\xi_k^-}\!\right)\\
G^{0,R}_{\downarrow\downarrow}(k,\epsilon)&=&\frac{1}{2\Delta_k}\!\left(\frac{\Delta_k-v_0k}{\epsilon\!+i0^+\!-\xi_k^+}+\frac{\Delta_k+v_0k}{\epsilon\!+i0^+\!-\xi_k^-}\!\right)\\
G^{0,R}_{\uparrow\downarrow}(k,\epsilon)&=&\frac{\eta\alpha}{2\Delta_k}\!\left(\frac{1}{\epsilon\!+i0^+\!-\xi_k^+}-\frac{1}{\epsilon\!+i0^+\!-\xi_k^-}\!\right)\\
G^{0,R}_{\downarrow\uparrow}(k,\epsilon)&=&\frac{\eta\alpha^*}{2\Delta_k}\!\left(\frac{1}{\epsilon\!+i0^+\!-\xi_k^+}-\frac{1}{\epsilon\!+i0^+\!-\xi_k^-}\!\right)
\end{eqnarray}
where $\xi_k^\pm=\frac{k^2}{2m}-\mu\pm\Delta_k$ is the dispersion of upper band and lower band, 2$\Delta_k=2\sqrt{(v_0k)^2+\eta^2|\alpha|^2}$ is the energy splitting between two bands, $v_0=k_0/m$ is "recoil" velocity. The advanced Green's function is $G^A(k,\epsilon)=(G^R(k,\epsilon))^\dag$. Meanwhile, as we know $k$ is conserved, therefore $G^{K,A,R}(r_1,r_2)=G^{K,A,R}(r_1-r_2)$. For Keldysh Green's function, we have $G^K(k,\epsilon)=G^R(k,\epsilon)F(k,\epsilon)-F(k,\epsilon)G^A(k,\epsilon)$ where $F(k,\epsilon)$ is the distribution function. If the system reaches thermal equilibrium,  $F(k,\epsilon)=(1-2n_F(\epsilon)){\mathbbm 1}$ ($\mathbbm 1$ is a unit matrix in spin space, $n_F(\epsilon)=1/(\exp((\epsilon-\mu)/T)+1)$ is Fermi distribution at temperature T). However, steady state distribution $F(k,\epsilon)$ does not necessarily be thermal distribution, therefore in principle $F(k,\epsilon)$ need to calculated self-consistently. But here, for simplicity, we assume the fermion distribution is fixed at zero temperature Fermi-Dirac distribution. Our mission in this article is to obtain the excitation spectrum function ${\cal A}(k,\epsilon)=-\frac{1}{\pi}{\rm Im} G^R(k,\epsilon)$.

Feynman rules for Feynman diagrams in Keldysh formalism are introduced in Fig.~\ref{Feyn} based on our Keldysh path integral formalism.


\subsection{Dyson equations}

\begin{figure}[t]
\includegraphics[width=8.5cm]{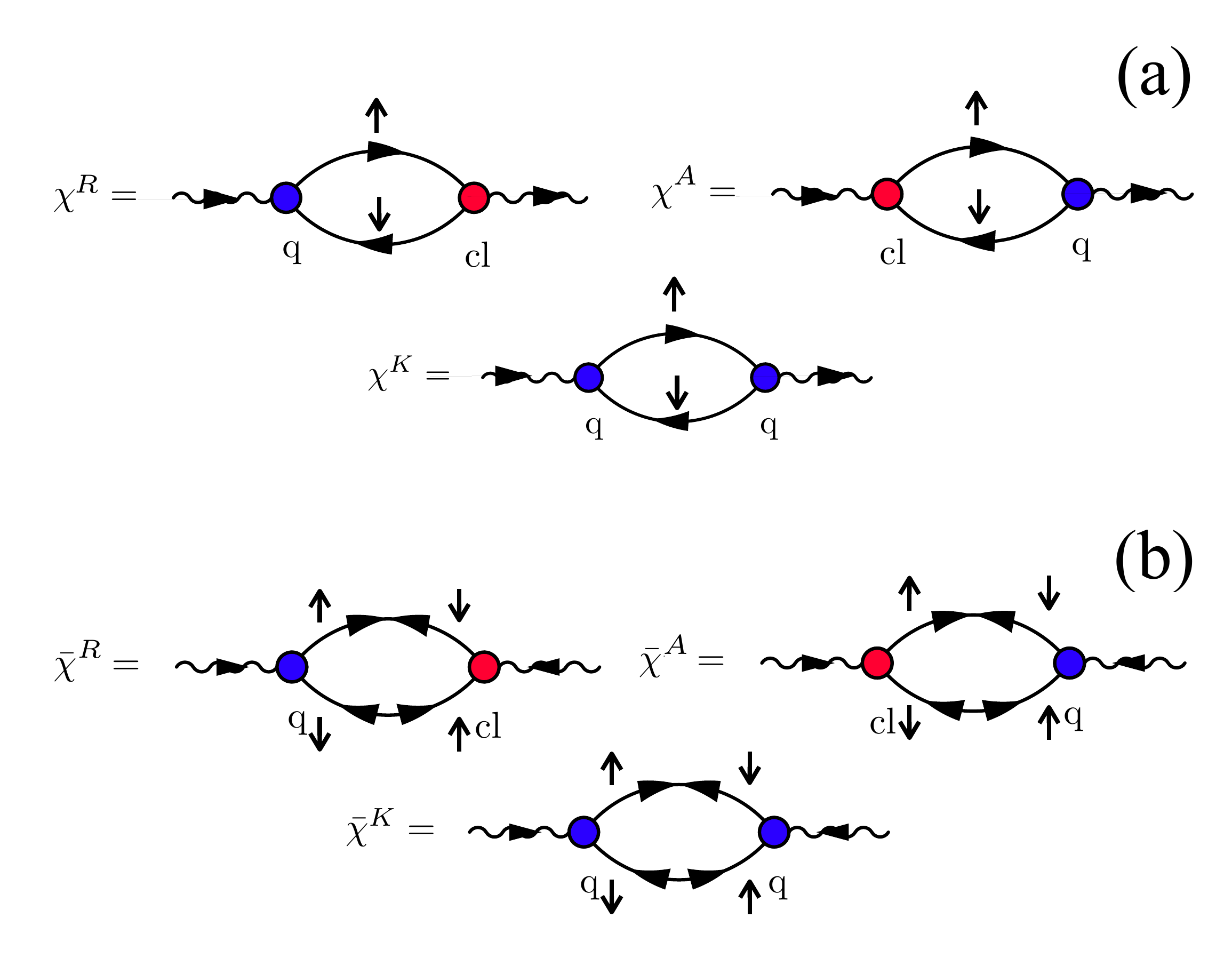}
\caption{Susceptibilities of fermions, also as cavity photon self-energy. (a) The normal photon self-energy contributions $\chi$. (b) The abnormal photon self-energy $\bar{\chi}$, which are nonzero only when $\alpha\neq0$. 
Pay attention that $G^{11}=G^R$, $G^{12}=G^K$, $G^{22}=G^A$ and $G^{21}=0$. The expressions for these Feynman diagrams are given in Eqn. (\ref{Sus1}), (\ref{Sus2}), (\ref{Sus3}), (\ref{Sus4}), (\ref{Sus5}) and (\ref{Sus6}).}\label{PhotonSelf}
\end{figure}

In this section, we will present self-consistent Dyson equations under steady state approximation. First of all, let us consider cavity photon self-energies, and this self-energy correction comes from polarization of fermions which could be shown diagrammatically in Fig. \ref{PhotonSelf}. Their explicit expressions are 

\begin{eqnarray}
\chi^{K}(\tilde{q})\!&=&\!\frac{i\eta^2}{2}\sum_{\tilde{k}}\left(G^{R}_{\uparrow\uparrow}(\tilde{k}+\tilde{q})G^A_{\downarrow\downarrow}(\tilde{k})+G^{A}_{\uparrow\uparrow}(\tilde{k}+\tilde{q})G^R_{\downarrow\downarrow}(\tilde{k})\right.\nonumber\\
&&\left.+G^{K}_{\uparrow\uparrow}(\tilde{k}+\tilde{q}) G^K_{\downarrow\downarrow}(\tilde{k})\right)\label{Sus1}\\
\bar{\chi}^K(\tilde{q})\!&=&\!\frac{i\eta^2}{2}\sum_{\tilde{k}}\left(G^{R}_{\uparrow\downarrow}(\tilde{k}+\tilde{q})G^A_{\downarrow\uparrow}(\tilde{k})+G^{A}_{\uparrow\downarrow}(\tilde{k}+\tilde{q})G^R_{\downarrow\uparrow}(\tilde{k})\right.\nonumber\\
&&\left.+G^{K}_{\uparrow\downarrow}(\tilde{k}+\tilde{q}) G^K_{\downarrow\uparrow}(\tilde{k})\right)\label{Sus2}
\end{eqnarray}

\begin{eqnarray}
\!\!\!\!\!\!\!\!\chi^R(\tilde{q})\!&=&\!\!\!\frac{i\eta^2}{2}\!\sum_{\tilde{k}}G_{\downarrow\downarrow}^R(\tilde{k}\!+\!\tilde{q}) G^K_{\uparrow\uparrow}(\tilde{k})+G^K_{\downarrow\downarrow}(\tilde{k}+\tilde{q})G^A_{\uparrow\uparrow}(\tilde{k})\label{Sus3}\\
\!\!\!\!\!\!\!\!\bar{\chi}^R(\tilde{q})\!&=&\!\!\!\frac{i\eta^2}{2}\!\sum_{\tilde{k}}G_{\downarrow\uparrow}^R(\tilde{k}\!+\!\tilde{q}) G^K_{\downarrow\uparrow}(\tilde{k})+G^K_{\downarrow\uparrow}(\tilde{k}+\tilde{q})G^A_{\downarrow\uparrow}(\tilde{k})\label{Sus4}
\end{eqnarray}

\begin{eqnarray}
\!\!\!\!\!\!\!\!\chi^A(\tilde{q})\!&=&\!\!\!\frac{i\eta^2}{2}\!\sum_{\tilde{k}}G_{\downarrow\downarrow}^A(\tilde{k}\!+\!\tilde{q}) G^K_{\uparrow\uparrow}(\tilde{k})+G^K_{\downarrow\downarrow}(\tilde{k}+\tilde{q})G^R_{\uparrow\uparrow}(\tilde{k})\label{Sus5}\\
\!\!\!\!\!\!\!\!\bar{\chi}^A(\tilde{q})\!&=&\!\!\!\frac{i\eta^2}{2}\!\sum_{\tilde{k}}G_{\downarrow\uparrow}^A(\tilde{k}\!+\!\tilde{q}) G^K_{\downarrow\uparrow}(\tilde{k})+G^K_{\downarrow\uparrow}(\tilde{k}+\tilde{q})G^R_{\downarrow\uparrow}(\tilde{k})\label{Sus6}
\end{eqnarray}
where $\tilde{q}=(0,\omega)$, $\tilde{k}=(k,\epsilon)$.
Then let us define susceptibility matrices
\begin{eqnarray}
\hat{\chi}^{i=R,A,K}(\omega)\equiv\left(
\begin{tabular}{cc}
$\chi^{i}(\omega)$&$\bar{\chi}^i(\omega)$\\
$(\bar{\chi}(-\omega))^*$&$(\chi^{i}(-\omega))^*$
\end{tabular}
\right),
\end{eqnarray}
where we denote the space as time-reversal space, in short, T space. 
then the Dyson equations for cavity photons can be written as
\begin{eqnarray}
\left(\begin{tabular}{cc}
$\hat{\Pi}^K$&$\hat{\Pi}^R$\\
$\hat{\Pi}^A$&0
\end{tabular}
\right)^{-1}=\left(\begin{tabular}{cc}
$\hat{\pi}_0^K$&$\hat{\pi}_0^R$\\
$\hat{\pi}_0^A$&0
\end{tabular}
\right)^{-1}-\left(\begin{tabular}{cc}
$0$&$\hat{\chi}^A$\\
$\hat{\chi}^R$&$\hat{\chi}^K$
\end{tabular}
\right)\label{eq:PhSelf}
\end{eqnarray}
where $\hat{}$\ \  in $\hat{\Pi}$, $\hat{\pi}$, $\hat{\chi}$ means matrix in T space (time-reversal space).
Further, let us discuss the symmetry of $\Pi^R(\omega)$ in T space. $\Pi_{11}^R(\omega)=-i\langle a(\omega)a^\dag(\omega)\rangle$, $\Pi_{22}^R(\omega)=-i\langle a^\dag(-\omega)a(-\omega)\rangle$, then we have $\Pi_{11}(\omega)=-\Pi_{22}^*(-\omega)$.

\begin{figure}[t]
\includegraphics[width=8.5cm]{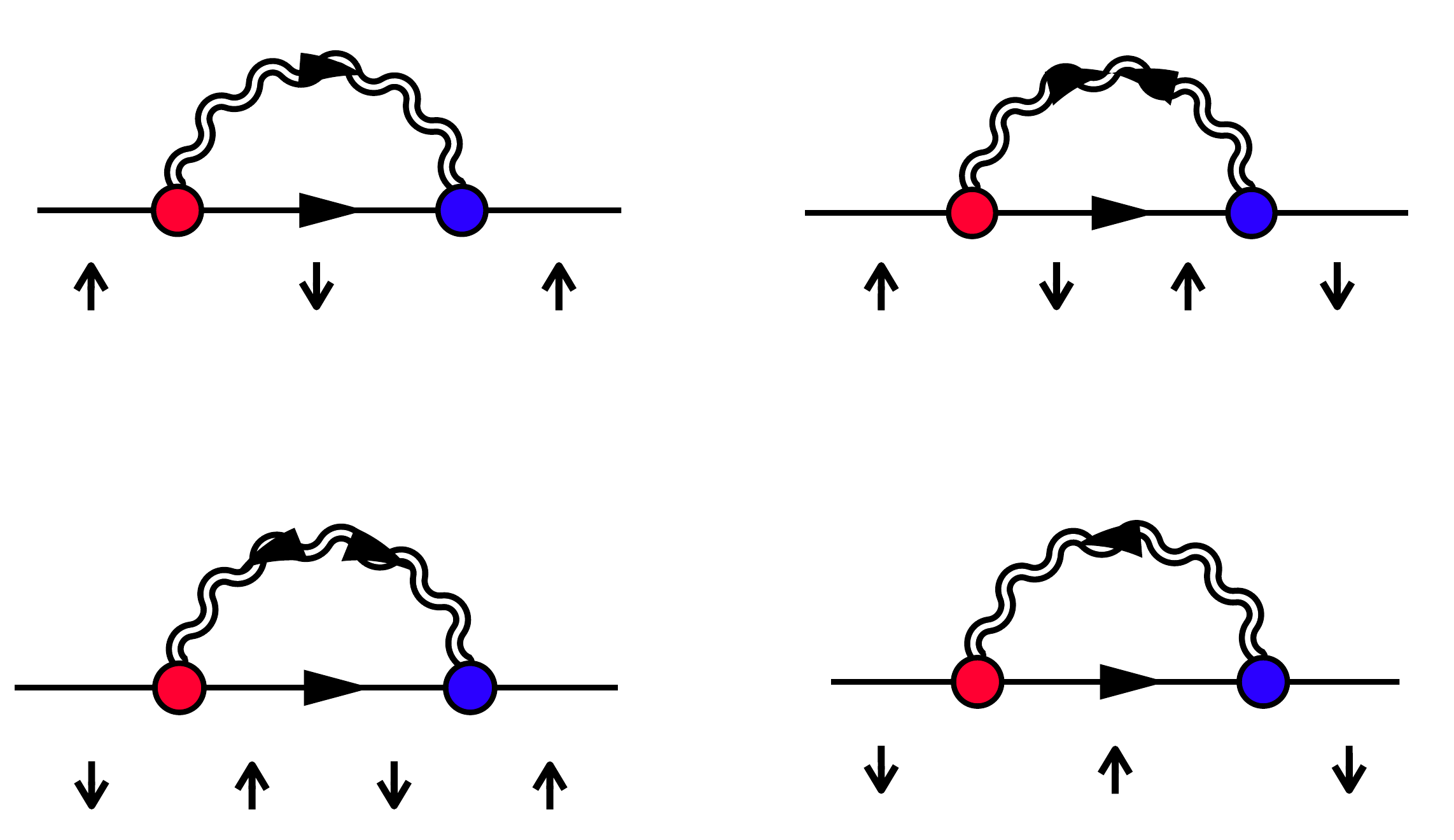}
\caption{Retard self-energy of fermions. Double wiggly line represents dressed cavity photon propagator $\hat{\Pi}(\omega)$. Self-energies expressions by these Feynman diagrams are given explicitly in Eqn. (\ref{eq: Self01}) }\label{FermionSelf}
\end{figure}

For the self-energy of fermions, we consider the Hatree term and the Fock terms. 
Since the Hatree terms can be written as $i\frac{\eta^2}{2}\Pi^K(0)\sum_{k,\epsilon} (G^R(k,\epsilon)+G^A(k,\epsilon))$, this is a constant shift of the chemical potential. In our scheme, we will fix the chemical potential of fermions, therefore we add the Hatree term contribution to chemical potential and require this dressed chemical potential to be fixed. On the other hand, the Fock terms can be shown diagrammatically as Fig. \ref{FermionSelf}. These self-energies can be explicitly written down as
\begin{eqnarray}
\!\Sigma^{\rm R}_{\sigma\sigma'}(\epsilon)\!&=&\!\!\!\frac{i\eta^2}{2}\int \!\!\frac{d\omega}{2\pi}\left( G^{\rm R}_{\bar{\sigma}\bar{\sigma}'}(\epsilon\!-\!\omega)\Pi_{\bar{\sigma}\bar{\sigma}'}^{\rm K}\!(\omega)\right.\nonumber\\
&&\hspace{1cm}\left.+ G^{\rm K}_{\bar{\sigma}\bar{\sigma}'}(\epsilon\!-\!\omega)\Pi_{\bar{\sigma}\bar{\sigma}'}^{\rm A}\!(\omega)\right)\label{eq: Self01}
\end{eqnarray}
\begin{eqnarray}
\!\Sigma^{\rm A}_{\sigma\sigma'}(\epsilon)\!&=&\!\!\!\frac{i\eta^2}{2}\int \!\!\frac{d\omega}{2\pi}\left( G^{\rm A}_{\bar{\sigma}\bar{\sigma}'}(\epsilon\!-\!\omega)\Pi_{\bar{\sigma}\bar{\sigma}'}^{\rm K}\!(\omega)\right.\nonumber\\
&&\hspace{1.2cm}\left.+ G^{\rm K}_{\bar{\sigma}\bar{\sigma}'}(\epsilon\!-\!\omega)\Pi_{\bar{\sigma}\bar{\sigma}'}^{\rm R}\!(\omega)\right)\label{eq: Self02}
\end{eqnarray}
\begin{eqnarray}
\!\Sigma^{\rm K}_{\sigma\sigma'}(\epsilon)\!&=&\!\!\!\frac{i\eta^2}{2}\int\frac{d\omega}{2\pi}\left[G^{\rm K}_{\bar{\sigma}\bar{\sigma}'}(\epsilon\!-\!\omega)\Pi_{\bar{\sigma}\bar{\sigma}'}^{\rm K}\!(\omega)+ \right.\nonumber\\
\!&&\!\!\!\left.G^{\rm R}_{\bar{\sigma}\bar{\sigma}'}(\epsilon\!-\!\omega)\Pi_{\sigma\sigma'}^{\rm A}\!(\omega)\!+\!G^{\rm A}_{\bar{\sigma}\bar{\sigma}'}(\epsilon\!-\!\omega)\Pi_{\bar{\sigma}\bar{\sigma}'}^{\rm R}\!(\omega)\right]\label{eq: Self03}
\end{eqnarray}
where $\sigma,\sigma'=1,2$ are spin index in $\hat{\Sigma}$, and $\hat{G}$,  or 1,2 in T space of $\hat{\Pi}$. $\bar{\sigma}$ means exchange $1$ and $2$, for example, $\bar{1}=2$, $\bar{2}=1$.
Finally, we get Dyson's equations for fermion's Green's functions as
\begin{eqnarray}
\left(\begin{tabular}{cc}
$\hat{G}^R$&$\hat{G}^K$\\
$0$&$\hat{G}^A$
\end{tabular}
\right)^{-1}\!\!\!\!\!\!\!(k,\epsilon)\!=\!\left[\left(\begin{tabular}{cc}
$\hat{G}_0^R$&$\hat{G}_0^K$\\
$0$&$\hat{G}_0^A$
\end{tabular}
\right)^{-1}\!\!\!\!\!-\left(\begin{tabular}{cc}
$\hat{\Sigma}^R$&$\hat{\Sigma}^K$\\
$0$&$\hat{\Sigma}^A$
\end{tabular}
\right)\right]\!\!(k,\epsilon)
\end{eqnarray}

One could always apply a rotation in spin space at every k to make the imaginary part of $G^R$ ( in other words, the spectrum) to be diagonalized. 
We denote the rotated Green's functions as $G^R_\pm(k,\epsilon)$, $G^A_\pm(k,\epsilon)$ and $G^K_\pm(k,\epsilon)$, where $\pm$ are band indices.
The fermionic excitation spectrum can be given as
 \begin{eqnarray}
 {\cal A}_{\pm}(k,\epsilon)=-\frac{1}{\pi}{\rm Im} G_{\pm}^R(k,\epsilon),
 \end{eqnarray}

Within steady state assumption, Eqn. (\ref{eq: Self01}), Eqn. (\ref{eq: Self02}), Eqn. (\ref{eq: Self03}) and Eqn.(\ref{eq:PhSelf}) formed a set of self-consistent equations for spectrum of fermions and cavity field, as well as distributions of fermions and cavity field. 
In this article, we will focus on the spectrum and we assume the fermion distribution is still fixed at zero temperature Fermi-Dirac distribution at the first self-consistent loop level.



\section{Numerical Results for Single Particle Exciation Spectrum}

In this section, we will show the numerical results for cavity photon spectrum and fermion excitation spectrum in different parameter regions. We will finally answer the question that when will the quasi-particle picture be valid.





\subsection{Single particle excitation spectrum of fermions}

Before we present the numerical results for fermions' spectrum function, let us first analyze when will the spectrum deviate from sharp quasi-particle peaks, or in other words, when will the self-energy $\hat{\Sigma}$ get large corrections. By Eqn. (\ref{eq: Self01}), (\ref{eq: Self02}), (\ref{eq: Self03}), we could learn that both cavity,  fermion spectrum and distribution contribute to the self-energy. Since the cavity photons are in a steady state rather than an equilibrium state, the typical distribution function is $\hat{F}_B(\omega)=\mathbbm{1}+2\hat{n}_B(\omega)={\rm diag}\{1,-1\}$, which is quite different from the zero temperature equilibrium distribution function $F_{B}^{\rm eq}(\omega)={\rm sgn}(\omega)$. Considering dressed photon will not change this steady state distribution much, we will use this approximate steady state distribution for qualitative estimation. At the same time,  let us make use of $\hat{G}^{R}_0$ and $\hat{G}^K_0$'s imaginary parts being delta functions, then we can get simplified expressions for fermion self-energies as 
\begin{eqnarray}
\!\!{\rm Im}\Sigma_{\uparrow\uparrow}^R(\epsilon_\pm)\!\!&\approx&\!\!i\eta^2P^\mp_{\downarrow\downarrow}(n_B^{22}(-\Delta \epsilon)\!+\!n_F(\epsilon_\mp)){\rm Im}\Pi^R_{22}(-\Delta\epsilon)\nonumber\\
\!\!{\rm Im}\Sigma_{\downarrow\downarrow}^R(\epsilon_\pm)\!\!&\approx&\!\!i\eta^2P^\mp_{\uparrow\uparrow}(n_B^{11}(\Delta \epsilon)\!+\!n_F(\epsilon_\mp)){\rm Im}\Pi^R_{11}(\Delta\epsilon)
\end{eqnarray}
where $\rm Im$ stands for imaginary part, $\Delta \epsilon=\epsilon_+-\epsilon_-$ is the energy gap between two bands, $P_{\uparrow\uparrow}^\pm=\frac{1}{2}(1\pm v_0k/\Delta_k)$, $P_{\downarrow\downarrow}^{\pm}=\frac{1}{2}(1\mp v_0k/\Delta_k)$ are spin projection factors. $n_{B}^{11}(\omega)$ and $n_{B}^{22}(\omega)$ are cavity photon steady state distribution functions, $1,2$ are T space indices. From our calculation we found $n_B^{11}(\omega)=0$ and $n_B^{22}(\omega)=-1$ for almost every $\omega$. Therefore we could further simplify the expressions as
\begin{eqnarray}
\!\!{\rm Im}\Sigma_{\uparrow\uparrow}^R(\epsilon_\pm)\!\!&\approx&\!\!-i\eta^2P^\mp_{\downarrow\downarrow}(1\!-\!n_F(\epsilon_\mp)){\rm Im}\Pi^R_{22}(-\Delta\epsilon)\nonumber\\
\!\!{\rm Im}\Sigma_{\downarrow\downarrow}^R(\epsilon_\pm)\!\!&\approx&\!\!i\eta^2P^\mp_{\uparrow\uparrow} n_F(\epsilon_\mp){\rm Im}\Pi^R_{11}(\Delta\epsilon)
\end{eqnarray}
From above expression we could see large  self-energy correction for spin up branch is only possible when $n_F(\epsilon_\pm)=0$, that is k mode being not occupied. A large correction for spin down branch is only possible when $n_F(\epsilon_\pm)=1$. Bare in mind that spin up branch shifts to the left and spin down branch shift to the right side, then we can check our claim in spectrum function visually.  

\begin{figure}
\includegraphics[width=8.cm]{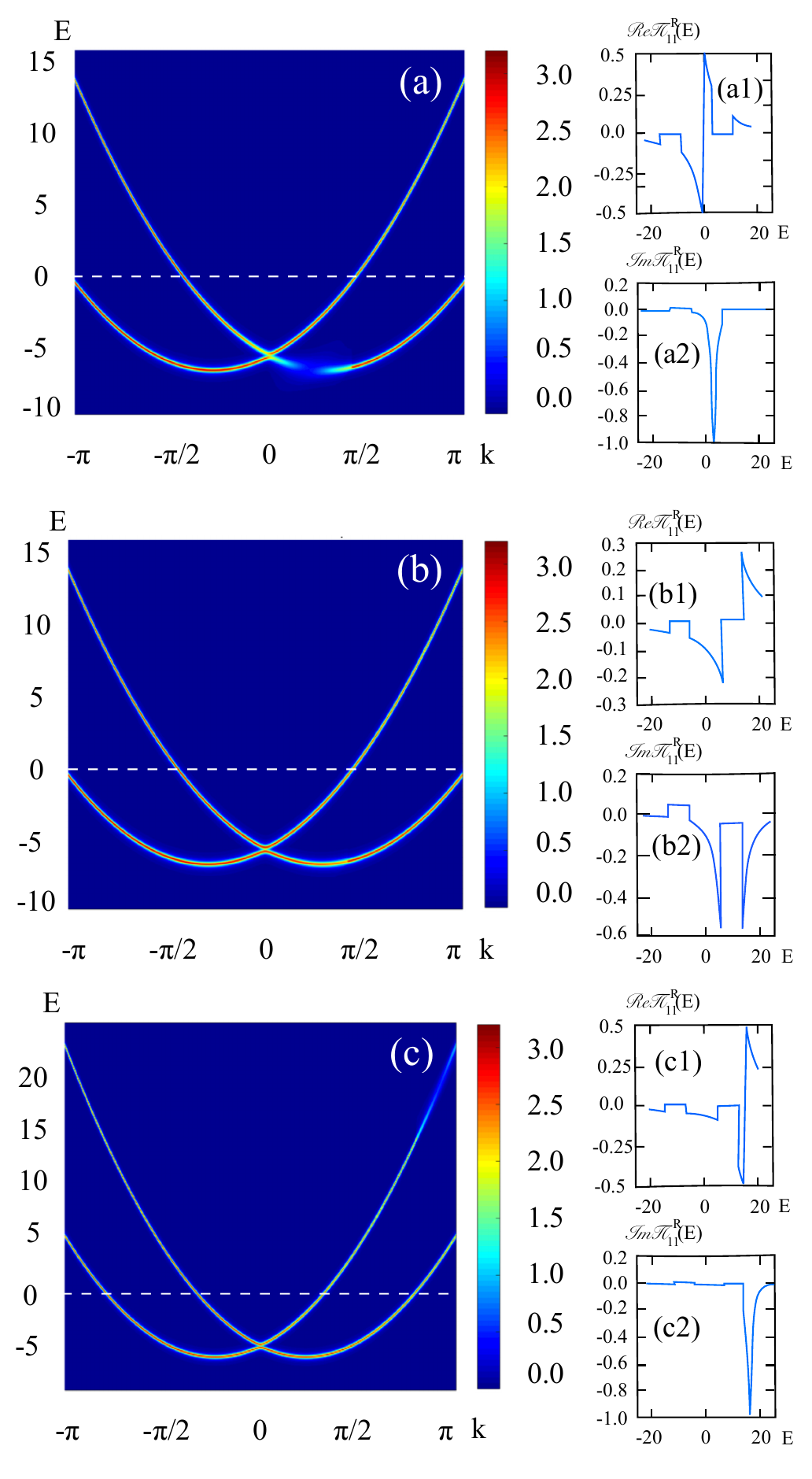}

\caption{Fermion spectrum at high filling factor. In all three figures, cavity decay rate $\kappa=E_r$, cavity detune is $\delta_c=-70E_r$, and the pumping field strength is $\epsilon_p=80E_r$, the atom number is $1600$, chemical potential is $\mu=6E_r$. The chemical potential is taken to be $E=0$ and labeled as white dashed lines in three figures. 
In (a), $\eta=0.78E_r$, $\cal A(\epsilon)$ is shown for the cavity excitation pole being in double occupancy region while in (b), ($\eta=0.738E_r$) $\cal A(\epsilon)$ is shown for the cavity excitation pole being in single occupancy region and finally, while finally in (c) ($\eta=0.7E_r$), $\cal A(\epsilon)$ is shown for the cavity excitation pole being in zero occupancy region. In (a1),(a2), (b1), (b2) and (c1), (c2) we give the real part and imaginary part of cavity photon's Green's function. The white dashed line is the fermion level which we take as zero energy.}\label{High}
\end{figure}

The second important ingredient  for large $\Sigma$ is big ${\rm Im}\Pi^R_{\sigma\sigma'}(\Delta \epsilon)$. In this part we discuss the condition for large ${\rm Im}\Pi^R_{\sigma\sigma'}(\pm2\Delta_k)$ where $2\Delta_k$ is the band gap at momentum k. As $[\Pi^{-1}]^R=\omega+\delta_c+i\kappa-\chi^R$, therefore large ${\rm Im}\Pi^R_{\sigma\sigma'}(\pm\Delta_k)$ means the real part and imaginary part of $[\Pi^{-1}]^R$ being both small. First of all, we could always find a momentum k so that the real part of $\Pi$ is small, thus we take $k^*$ to denote the momentum for smallest ${\rm Re}[\Pi^{-1}]^R$. On the other hand, the imaginary part of $[\Pi^{-1}]^R$ is determined by the cavity decay rate $\kappa$ and density of states (DOS) of particle hole excitations ${\rm Im}\chi^R$, therefore large $\Pi^R$ requires small $\kappa$ and vanishing ${\rm Im}\chi^R(\omega=\pm2\Delta_k)$. Small $\kappa$ requires high finesse cavity while vanishing ${\rm Im}\chi^R(\omega=\pm2\Delta_k)$ means no particle-hole (PH) excitation at momentum k. Because the atomic gas and cavity photons are in strong interacting region therefore as long as ${\rm Im}\chi^R(\omega=\pm2\Delta_k)$ exists, it is large ($\sim $ MHz $\gg$ recoil energy). Further, DOS of PH excitations is nonzero only for single occupancy when $n_F(\epsilon_-)=1$ and $n_F(\epsilon^+)=0$, therefore small $[\Pi^{-1}]^R$ could be only achieved in double occupied region and zero occupied region. 

In the following paragraph, we will first present our data for different typical filling and pole positions of polariton for $\kappa\sim E_r$, then we will go on to present how $\kappa$ changes the spectrum broadening.

\begin{figure}[t]
\includegraphics[width=7.9cm]{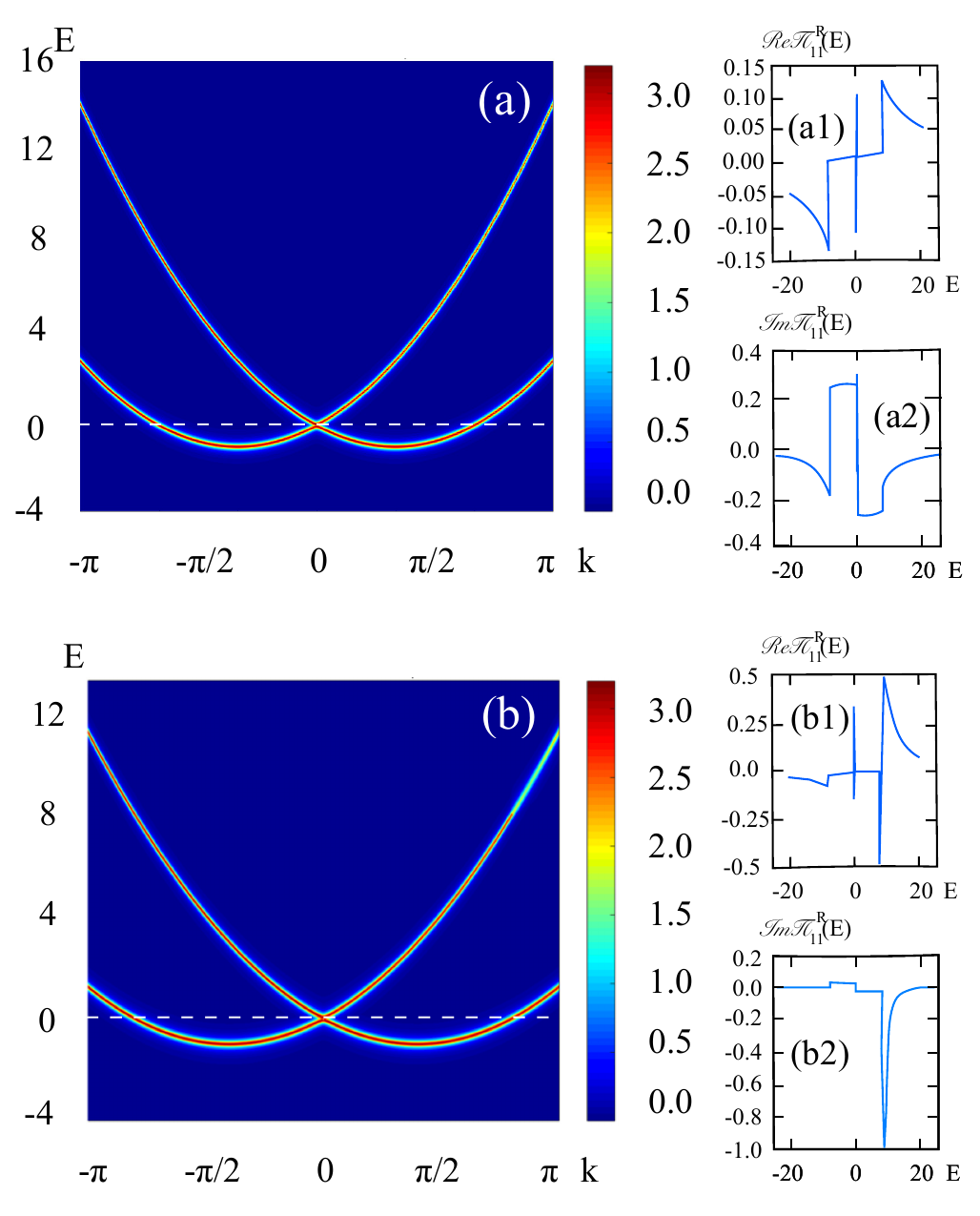}
\caption{Fermion spectrum at middle filling when the Fermi level pass through the band gap. $\mu=E_r$ with other parameters except $\eta$ unchanged. In (a) ($\eta=0.31E_r$), the cavity photon excitation energy matches $2\Delta_k$ for $k$ in single filling region. In (b) ($\eta=0.29E_r$), the cavity photon excitation energy matches $2\Delta_k$ for a $k$ in zero filling region.}\label{Middle}
\end{figure}

\begin{figure}[ht]
\includegraphics[width=8.cm]{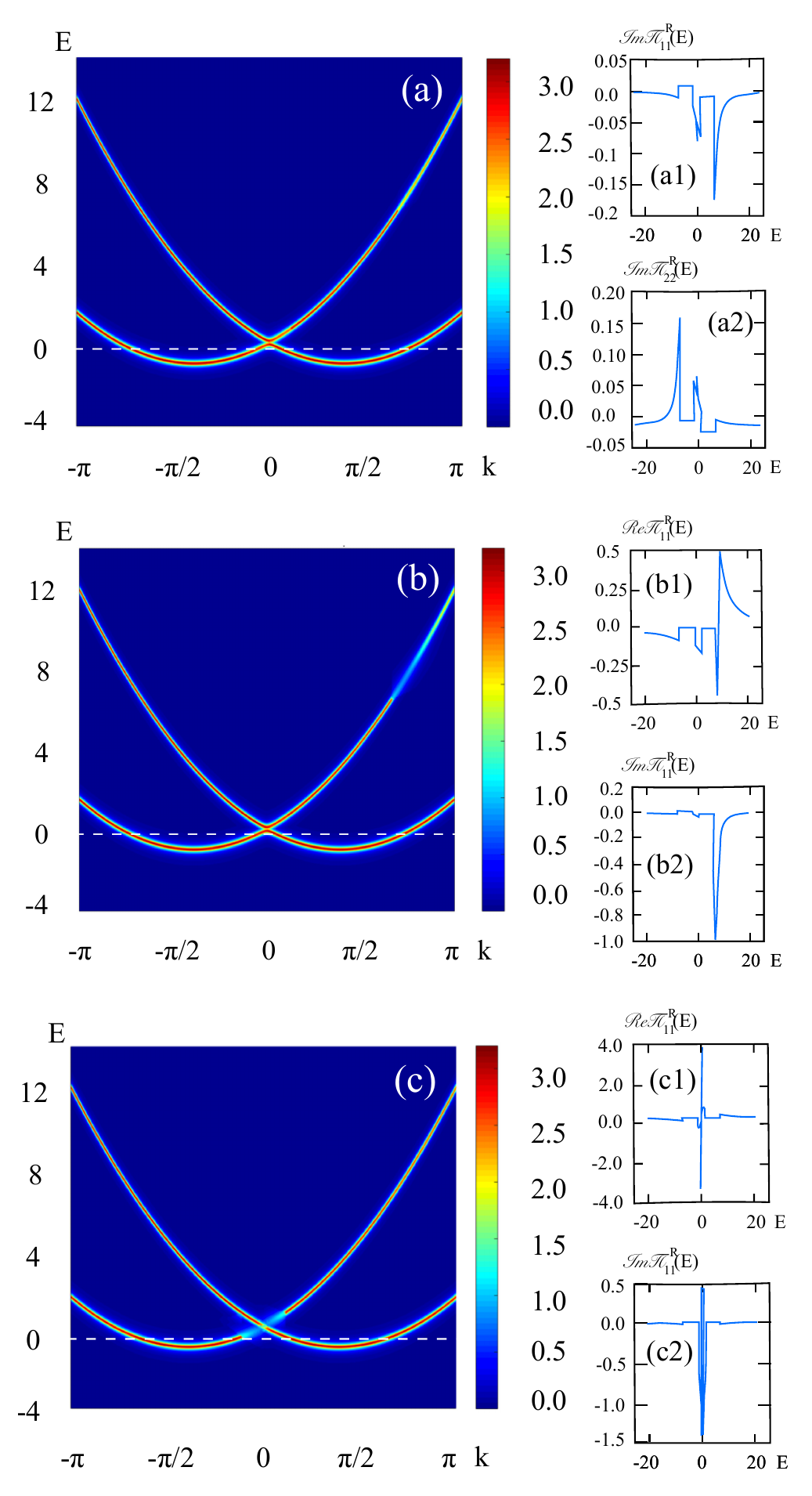}
\caption{Fermion spectrum at low filling when the Fermi level is below the band gap. $\mu=0.5E_r$ and other parameters except $\eta$ are unchanged. In (a) ($\eta=0.537E_r$), the cavity photon excitation energy matches $2\Delta_k$ for $k$ in single filling region. In (b) ($\eta=0.525E_r$), the cavity photon excitation energy matches $2\Delta_k$ for a $k$ in zero filling region for large $k$. In (c) ($\eta=0.556E_r$), the cavity photon excitation energy matches $2\Delta_k$ for a $k$ in zero filling region for small $k$}
\label{Low}
\end{figure}

In high filling case where double occupancy is possible, there are three typical cases, determined by polariton pole position determined by ${\rm Re}[\Pi^R_{11}(\omega^*)]^{-1}=0$, whose solution we denoted as $\omega^*$. $\omega^*$ will match certain $2\Delta_{k^*}$ for a specified $k^*$ (If $\omega^*<2\Delta_{k=0}=2\eta|\alpha|$, we take $k^*=0$). The position of $k^*$ can be classified by how k mode is filled. First suppose $0<k_{F1}<k_{F2}$ are two real solution for $\xi_{\pm}(k)=0$, then the first typical case is $|k^*|\leq k_{F1}$, we call double filling region because in this region both upper band and lower band are occupied; the second case is $k_{F1}<|k^*|\leq k_{F2}$, which we addressed as single filling region and finally $|k^*|>k_{F2}$, zero filling region.
Both the fermion excitation spectrum and photon excitation spectrum are given in Fig. \ref{High}(a), (b), (c). One could see only in (a) and (c) cases, fermion excitation spectrum have large corrections. As we addressed previously, a large correction in self-energy $\hat{\Sigma}$ requires large $\Pi^R$ satisfying $1-n_F(\epsilon_{k^*}^-)\sim 1$ or $n_F(\epsilon_{k^*}^+)\sim 1$. In case (a), $n_F(\epsilon_{k^*}^+)\sim 1$ is satisfied, a large correction could be expected in spin down branch when we get large density of cavity excitation photons. The result is shown in Fig. \ref{High}(a), and in (a1) and (a2) we show real part and imaginary part of $\Pi^R_{11}(\omega)$. To summarize the result, first, the $Z_2$ symmetry of the spectrum is broken by dissipation, only one branch is unstable by the choice of dissipation. Second, it is the lowest energy of the whole spectrum that is modified, leading to a vacuum change for fermions. This is the most non-trivial case where we could reach a different vacuum due to dissipation. In the second case Fig. \ref{High}(b), a small $\Pi^R_{11}(\omega^*)$ is expected as the imaginary part of $[\Pi^R(\omega)]^{-1}$ is dominated by large particle-hole excitation DOS, a dissipation channel opened by atom-light interaction. Also because of the filling selection, the self-energy $\hat{\Sigma}$ gets very small corrections, therefore the spectrum remains a prefect quasi-particle spectrum. We also calculated the distribution function of fermions in this region by accessing $\hat{G}^K$, where we found $n_F(\epsilon)$ is close to zero temperature fermi distribution as we initially supposed. Therefore in this region, the spectrum is almost unchanged, one can safely omit the corrections from dissipation. Finally, when $|k^*|>k_{F2}$, the "spin up" band get large self-energy correction. Because the low energy spectrum is unchanged in this region, this spectrum variation is secondary and will only change the details in "high" temperature physics. Still, the reason for symmetry broken is the choice of $1-n_F(\epsilon_{k^*}^-)\sim 1$.

Results of moderate filling (chemical potential in band gap) and shallow filling are presented in Fig. \ref{Middle} and Fig. \ref{Low} respectively. From previous analysis, we can find when the chemical potential is right in the gap, the particle hole excitation DOS is always large, therefore the quasi particle pole is not changed. This is verified by Fig. \ref{Middle} (a) and (b). The spectrum broadening in spin up branch of Fig. \ref{Middle}(b) is caused by zero occupancy for k mode. Finally, we could also analyze shallow filling case while an extra region is possible as is shown in Fig. \ref{Low}(c). When the cavity photon excitation gap is very small, then the spin up spectrum is broadened for $|k|<k_{F1}$. But for all the cases except the example given in Fig. \ref{High}(a), the spectrum broadening is not around lowest energy state, therefore less important. In the next section, we will focus on case Fig. \ref{High}(a), and give the cavity decay rate dependence on the maximal broadening of the spectrum.



\subsection{Spectrum dependence on cavity decay rate}

Here in this section we will present how the maximal spectrum broadening changes against cavity decay rate $\kappa$. With an extremely small $\kappa$, the cavity becomes perfect, no photon leaks out, then the spectrum of fermions must be unchanged. For a small cavity decay rate $\kappa\ll E_r$, when particle-hole excitations are suppressed, we find the fermion excitation spectrum will split into two peaks, whose distance increases with $\kappa$. For $\kappa\sim E_r$, two peaks will merge into one peak and the spectrum width will reach its maximum. Further increasing of cavity decay rate will lead to a strong coupling between environment and system, large dissipation like a frequent measurement could pin the system in its original spectrum. This is also the situation we find numerically, the quasi particle peak becomes sharper and sharper when $\kappa$ becomes much larger than recoil energy.

\begin{figure}[h]
\includegraphics[width=8.2cm]{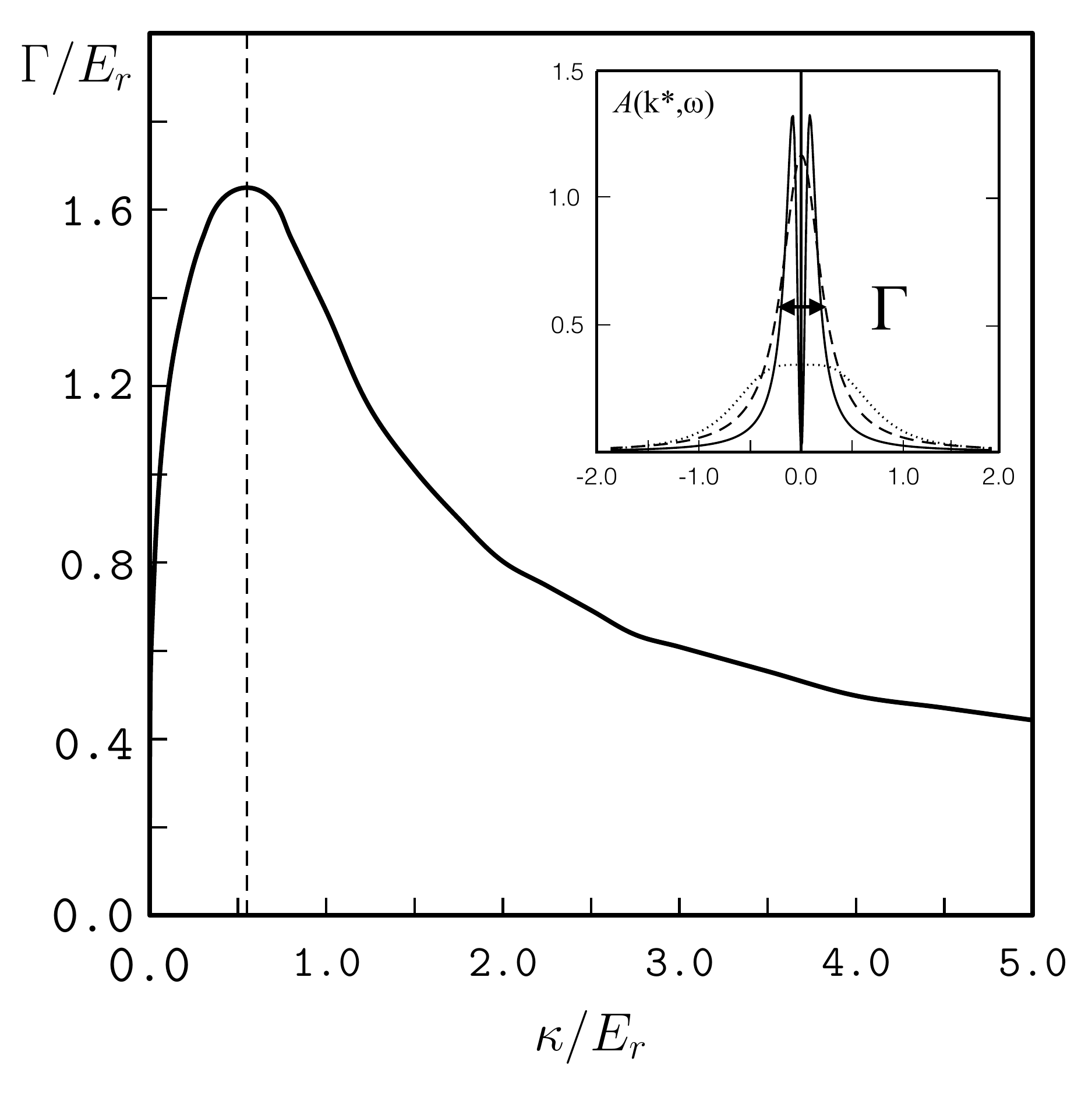}
\caption{ Here we define $\Gamma$ to be the full width of half maximum of ${\cal A}(k^*,\omega)$. Three typical ${\cal A}(k^*,\omega)$ for $\kappa\ll E_r$ (solid line), $\kappa\sim E_r$ (dotted line) and $\kappa\gg E_r$ (dashed line) are given in the inner figure. $\Gamma$ is found to be peaked at $\kappa\sim 0.5E_r$ }\label{Broadening}
\end{figure}

In Fig. \ref{Broadening}, we show how this full width at half maximum of spectrum function changes against cavity decay rate $\kappa$. The full width at half maximum of spectrum ${\cal A}(k^*,\omega)$ is defined as $\Gamma$, displaying in the inner figure of Fig. \ref{Broadening}. One can observe that the spectrum at very small $\kappa$ split into two peaks, that is because for small $\kappa$, the resonance of cavity photon and atom gap is quite exact, but the decay process is blocked by Pauli principle. The splitting of the spectrum reconcile Pauli principle and the requirement of resonance which brings photon absorption process on shell.

\section{Conclusion}

Based on Keldysh formalism, we calculated the spectrum function ${\cal A}_{\pm}(k,\omega)=\frac{i}{2\pi}(G_{\pm}^R(k,\omega)-G_{\pm}^A(k,\omega))$ by obtain $G^{R(A)}_{\pm}(k,\omega)$ with the help of Dyson equations. We found that the cavity dissipation will result a broadening when the band gap resonant with polariton excitation energy, and this broadening becomes large when the cavity decay rate $\kappa$ matches recoil energy. Interestingly enough, this broadening is highly spin selective depending on how fermion is filled and make the spectrum asymmetric. To be specific, there will be no obvious spectrum broadening when the momentum $k^*$ for resonant gap $2\Delta_{k^*}$ is singly occupied; there will be a spectrum broadening on spin down branch for $k^*$ being double occupied and a broadening on spin up branch for $k^*$ being zero occupied. The most interesting case is when the chemical potential allows double occupancy and the low energy spectrum is affected, causing inconsistency against our ground state assumption. Further study on this special parameter region with a self-consistent treatment is needed in the future.

\section{Aknowledgement}

We thank Hui Zhai for motivating and encouraging us on this subject. We also thank Pengfei Zhang for helpful discussions. This project is supported by NSFC under Grants No. 11604225 and Foundation of Beijing Education Committees under Grants No. KM201710028004.

\newpage
\begin{widetext}
\appendix
\section{Analytical Expression for Dynamical Susceptibility}

Let us rewrite the retard $\chi^R$ in more explicit form,
\begin{eqnarray}
\chi^R(\omega)&=&\frac{i\eta^2}{2}\sum_{k,\epsilon}\left(G^R_{\uparrow\uparrow}(\epsilon+\omega, k)G^K_{\downarrow\downarrow}(\epsilon,k)+G^K_{\uparrow\uparrow}(\epsilon+\omega, k)G^A_{\downarrow\downarrow}(\epsilon,k)\right)\nonumber\\
&=&\frac{\eta^2}{2}\sum_{k}\left(\frac{P_{\uparrow\uparrow}^+ P_{\downarrow\downarrow}^-(1-2n_F^-)}{\omega-2\Delta_k+i0^+}+\frac{P_{\uparrow\uparrow}^- P_{\downarrow\downarrow}^+(1-2n_F^+)}{\omega+2\Delta_k+i0^+}+\frac{P_{\uparrow\uparrow}^+ P_{\downarrow\downarrow}^-(1-2n_F^-)}{-\omega-2\Delta_k-i0^+}+\frac{P_{\uparrow\uparrow}^- P_{\downarrow\downarrow}^+(1-2n_F^+)}{-\omega+2\Delta_k-i0^+}\right)\nonumber\\
&=&\chi_0(\omega+i0^+)+2\chi_1(\omega+i0^+)+\chi_2(\omega+i0^+)
\end{eqnarray} 
$P_{\uparrow\uparrow}^\pm=\frac{1}{2}(1\pm v_0k/\Delta_k)$, $P_{\downarrow\downarrow}^{\pm}=\frac{1}{2}(1\mp v_0k/\Delta_k)$ are projection factors, and $\chi_0$, $\chi_1$ and $\chi_2$ are introduced as follows
\begin{eqnarray}
\chi_0(\omega+i0^+)=\eta^2\sum_k\left[\frac{1}{4(\omega+2\Delta_k+i0^+)}\!-\!\frac{1}{4(\omega-2\Delta_k+i0^+)}\right]\!(n_{F,k}^-\!-\!n_{F,k}^+)\\
\chi_1(\omega+i0^+)=\eta^2\sum_k\left[\frac{d_z}{4(\omega-2\Delta_k+i0^+)}\!+\!\frac{d_z}{4(\omega+2\Delta_k+i0^+)}\right]\!(n_{F,k}^-\!-\!n_{F,k}^+)\\
\chi_2(\omega+i0^+)=\eta^2\sum_k\left[\frac{d_z^2}{4(\omega+2\Delta_k+i0^+)}\!-\!\frac{d_z^2}{4(\omega-2\Delta_k+i0^+)}\right]\!(n_{F,k}^-\!-\!n_{F,k}^+)
\end{eqnarray}
Similarly, we can find $\bar{\chi}^R(\omega)=\chi_0(\omega+i0^+)-\chi_2(\omega+i0^+)$.
In the following we are going to calculate analytical expressions for $\chi_{0,1,2}$.

\begin{eqnarray}
\chi_0(\omega+i0^+)
&=&\left.\frac{N\eta^2}{2\pi v_0}\left[\frac{\lambda_\omega}{\sqrt{1-\lambda_\omega^2}}\left(-\arctan\left(\sqrt{\frac{1-\lambda_\omega}{1+\lambda_\omega}}\tan\frac{\theta}{2}\right)-\arctan\left(\sqrt{\frac{1+\lambda_\omega}{1-\lambda_\omega}}\tan\frac{\theta}{2}\right)\right)-\ln\frac{1+\sin\theta}{1-\sin\theta}\right]\right|_{\theta_I}^{\theta_F}
\end{eqnarray}

where $\lambda_\omega=(\omega+i0^+)/2\eta|\alpha|$, $\theta_F=\arctan(v_0k_F/\eta|\alpha|)$, $\theta_I=\arctan(v_0k_I/\eta|\alpha|)$.

\begin{eqnarray}
\chi_1(\omega+i0^+)=\eta^2\sum_k\frac{2\omega d_z}{\omega^2-\Delta_k^2}
=\frac{N\eta^2}{2\pi v_0}\left(\ln\left|\frac{\omega+i0^++\Delta_F}{\omega+i0^+-\Delta_F}\right|-\ln\left|\frac{\omega+i0^++\Delta_I}{\omega+i0^+-\Delta_I}\right|\right),
\end{eqnarray}

where $\Delta_F=2\sqrt{v_0^2k_F^2+\eta^2|\alpha|^2}$, $\Delta_I=2\sqrt{v_0^2k_I^2+\eta^2|\alpha|^2}$. $\theta_{F,I}$, $\Delta_{F,I}$ contain information of filling fraction as well as Fermi surface.

\begin{eqnarray}
\chi_2(\omega)=\eta^2\sum_k\frac{d_z^2\Delta_k}{\omega^2-4\Delta_k^2}=\sum_k\frac{\eta^2\Delta_k}{\omega^2-4\Delta_k^2}-\eta^4|\alpha|^2\sum_k\frac{1}{\Delta_k(\omega^2-4\Delta_k^2)}
\end{eqnarray}

\begin{eqnarray}
\chi_2(\omega+i0^+)=\left.\chi_0(\omega+i0^+)-\frac{N\eta^2}{2\pi v_0}\frac{1}{\lambda_\omega\sqrt{1-\lambda_\omega^2}}\left[\arctan\left(\sqrt{\frac{1-\lambda_\omega}{1+\lambda_\omega}}\tan\frac{\theta}{2}\right)-\arctan\left(\sqrt{\frac{1+\lambda_\omega}{1-\lambda_\omega}}\tan\frac{\theta}{2}\right)\right]\right|_{\theta=\theta_I}^{\theta_F}
\end{eqnarray}
All imaginary part could be get from analytical continuation.

\section{Cavity Fluctuation Spectrum}
In this section, we are going to present the details for calculation of cavity fluctuation correlation function, $\hat{\Pi}^{R,A,K}(\omega)$. $\Pi_{11}^{\alpha\beta}(t,t')=-i\theta(t-t')\langle\delta a_{\alpha}(t)\delta a^\dag_{\beta}(t')\rangle$, $\Pi_{22}^{\alpha\beta}(t,t')=-i\theta(t-t')\langle \delta a^\dag_{\alpha}(t)\delta a_{\beta}(t')\rangle$, $\Pi^R_{12}(t,t')=-i\theta(t-t')\langle \delta a_{\alpha}(t)\delta a_{\beta}(t')\rangle$, $\Pi^R_{21}(t,t')=-i\theta(t-t')\langle \delta a^\dag_{\alpha}(t)\delta a^\dag_{\beta}(t')\rangle$. ($\alpha,\beta={\rm cl)(\rm q}$) $\hat{\Pi}^R_{2\times2}=\hat{\Pi}_{2\times2}^{\rm cl,q}$, $\hat{\Pi}^A_{2\times2}=\hat{\Pi}_{2\times2}^{\rm q,cl}$, $\hat{\Pi}^K_{2\times2}=\hat{\Pi}_{2\times2}^{\rm cl,cl}$, $\hat{\Pi}^{\rm q,q}_{2\times2}=0$. In our article we suppose steady state could be reached in long time limit, therefore $\hat{\Pi}^{\alpha\beta}_{2\times2}(t,t')=\hat{\Pi}^{\alpha\beta}_{2\times2}(t-t')$, and $\hat{\Pi}^{\alpha\beta}(\omega)=\int d(t-t')e^{i\omega(t-t')}\hat{\Pi}^{\alpha\beta}_{2\times2}(t-t')$.

The Dyson equations for cavity field correlation functions can be explicitly written as 
\begin{eqnarray}
\Pi^R_{ab}(t,t')&=&\pi_{ab}^{0R}(t-t')+\sum_{c,d=1,2}\left(\pi_{ac}^{0R}\circ\chi_{cd}^{R}\circ\Pi^R_{db}\right)(t,t'),
\end{eqnarray} 
where the matrix indices are in $2\times2$ T (time-reversal) space. 
According to steady state approximation, after Fourier transformation, we get
\begin{eqnarray}
\Pi^R_{ab}(\omega)&=&\pi_{ab}^{0R}(\omega)+\sum_{c,d=1,2}\pi_{ac}^{0R}(\omega)\chi_{cd}^{R}(\omega)\Pi^R_{db}(\omega)
\end{eqnarray}

\begin{figure*}[h]
\includegraphics[width=13cm]{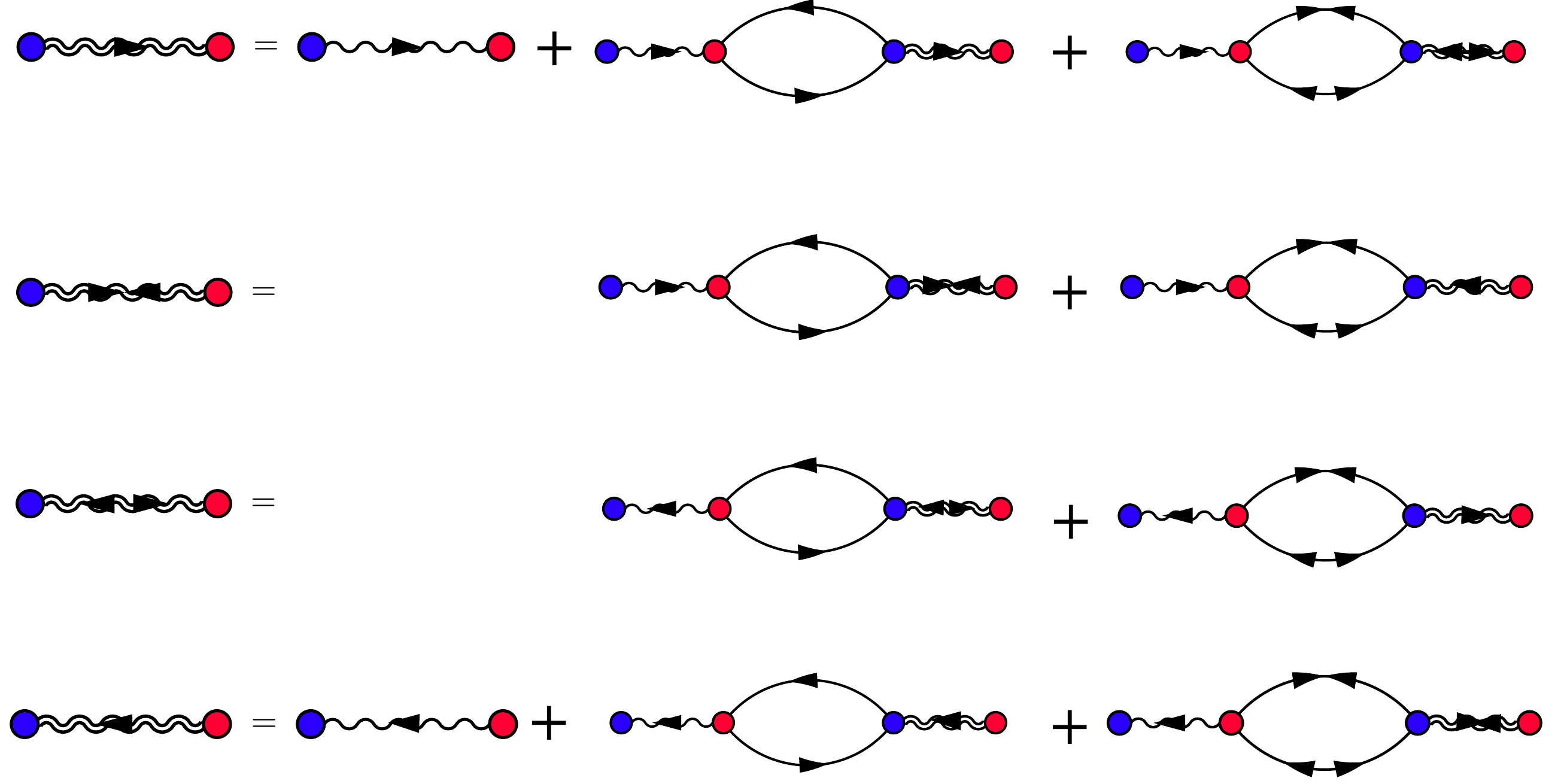}
\caption{Diagrammatic illustration of Dyson equation for $\Pi^R$. Here red dot represent ${\rm cl}$ vertex and blue dot represent ${\rm q}$ vertex. These Dyson equations are : (1) $\Pi^R_{11}=\pi^{0R}_{11}+\pi^{0R}_{11}\circ\chi_{11}^{R}\circ\Pi^R_{11}+\pi^{0R}_{11}\circ\chi_{12}^{R}\circ\Pi^R_{21}$; (2) $\Pi^R_{12}=\pi^{0R}_{11}\circ\chi_{11}^{R}\circ\Pi^R_{12}+\pi^{0R}_{11}\circ\chi_{12}^{R}\circ\Pi^R_{22}$; (3) $\Pi^R_{21}=\pi^{0R}_{22}\circ\chi_{22}^{R}\circ\Pi^R_{21}+\pi^{0R}_{22}\circ\chi_{21}^{R}\circ\Pi^R_{11}$; (4) $\Pi^R_{22}=\pi^{0R}_{22}\circ\chi_{22}^{R}\circ\Pi^R_{22}+\pi^{0R}_{22}\circ\chi_{21}^{R}\circ\Pi^R_{12}$. }\label{PiRDyson}
\end{figure*}
Finally, for the Keldysh Green's function, we have
\begin{eqnarray}
\Pi_{ab}^K(\omega)=\pi^{0K}_{ab}(\omega)+\sum_{c,d=1,2}\left(\pi_{ac}^{0K}(\omega)\chi_{cd}^{A}(\omega)\Pi^A_{db}(\omega)+\pi_{ac}^{0R}(\omega)\chi_{cd}^{R}(\omega)\Pi^K_{db}(\omega)+\pi_{ac}^{0R}(\omega)\chi_{cd}^{K}(\omega)\Pi^A_{db}(\omega)\right)
\end{eqnarray}




\end{widetext}

\end{document}